\documentclass[pra,twocolumn,preprintnumbers,amsmath,amssymb,superscriptaddress,showpacs,longbibliography]{revtex4-1}
\usepackage{graphicx,amsmath,amssymb,amsfonts,graphics,color,bm,dsfont}
\usepackage[mathscr]{eucal}

\newcommand{\ket}[1]    {| #1 \rangle}

\newcommand{\ketbra}[2]    {|#1\rangle\!\langle #2|}

\newcommand{\x}    {{\bf r}}

\newcommand{\on}{\text{On}}
\newcommand{\off}{\text{Off}}
\newcommand{\onstar}{{-}\text{On}}
\newcommand{\CHSHlike}{noise-averaged CHSH}
\newcommand{\CHSHlikeAD}{noise-averaged}

\usepackage{color}

\allowdisplaybreaks

\begin{document}

\author{J. Krzywda}
\affiliation{Institute of Physics, Polish Academy of Sciences, al.~Lotnik{\'o}w 32/46, PL 02-668 Warsaw, Poland}
\author{P. Sza\'{n}kowski}\email{piotr.szankowski@ifpan.edu.pl}
\affiliation{Institute of Physics, Polish Academy of Sciences, al.~Lotnik{\'o}w 32/46, PL 02-668 Warsaw, Poland}
\author{J. Chwede\'nczuk}
\affiliation{Faculty of Physics, University of Warsaw, ul. Pasteura 5, PL--02--093 Warsaw, Poland}
\author{{\L}. Cywi\'nski}
\email{lcyw@ifpan.edu.pl}
\affiliation{Institute of Physics, Polish Academy of Sciences, al.~Lotnik{\'o}w 32/46, PL 02-668 Warsaw, Poland}

\title{Decoherence-assisted detection of entanglement of two qubit states}

\begin{abstract}
  We show that the decoherence, which in the long run destroys  quantum features of a system, can be used to reveal the entanglement in a two-qubit system. 
  To this end,  we consider a criterion that formally resembles the Clauser-Horne-Shimony-Holt (CHSH) inequality. In our case
  the local observables are set by the coupling of each qubit to the environmental noise, controlled with the dynamical decoupling method.
  We demonstrate that the constructed inequality is an entanglement criterion---it can only be violated by non-separable initial two-qubit states, provided that the local noises are correlated.
  We also show that for a given initial state, this entanglement criterion can be repurposed as a method of discriminating between Gaussian and non-Gaussian noise generated by the environment of the qubits. The latter application is important for ongoing research on using qubits to characterize the dynamics of environment that perturbs them and causes their decoherence.
\end{abstract}
\maketitle

\section{Introduction}
The coupling of a system to the environment leads to decoherence, which is commonly viewed as an unfortunate but unavoidable obstacle to observing quantum 
effects~\cite{decoherence,Hornberger}. This is because the loss of coherence of a quantum system due to interaction with its surroundings is an important ingredient of the
the quantum-to-classical transition~\cite{Zurek_RMP03,joos2013decoherence}.
Therefore, to maintain the quantum features, the system must be highly isolated and thus protected against the decoherence. There are numerous ways for achieving this goal: 
using quantum error correction \cite{Shor_PRA95,Devitt_RPP13,Demkowicz_PRL14,Dur_PRL14,Kessler_PRL14,Arrad_PRL14}, 
restricting the dynamics to the decoherence-free subspace when noises experienced by multiple qubits are correlated \cite{Duan_PRA98,Jeske_NJP14}, 
or performing dynamical decoupling, i.e.,~subjecting the qubits to a sequence of unitary operations that make them less sensitive to the environmental noise
\cite{Viola_PRA98,Uhrig_PRL07,Gordon_PRL08,Cywinski_PRB08,Khodjasteh_NC13,Paz_NJP16,Suter_RMP16}.  Methods of harnessing the decoherence, dedicated for quantum metrology, have been proposed in various
scenarios \cite{Chin_PRL12,Chaves_PRL13,Smirne_PRL16,Szankowski_PRA14}, and in atomic systems coherence times are extended by fine-tuning of the inter-particle 
interactions~\cite{ferrari,geiger2018observation}.
With the advent of precise and customizable methods of qubit control it has become possible to embrace the environment and thus the process of decoherence.
Rather than treating it as an obstacle against the development of quantum technologies, the approach is to make it a part of a task, e.g.,
to characterize the environmental fluctuations by careful analysis of decoherence of a single qubit \cite{Degen_RMP17,Szankowski_JPCM17} and multiple qubits 
\cite{Szankowski_PRA16,Szankowski_JPCM17,Paz_PRA17,Krzywda_PRA17}, or even as a integral element of protocol for creation of  entanglement \cite{PhysRevLett.107.080503}.

Here we demonstrate that the paradigm of destructive decoherence can be reversed to some extent; we show that the coupling to the environment can become
a part of a protocol of detecting the entanglement between two qubits.
In more detail: we take two qubits, couple them to correlated sources of decoherence, 
and construct a criterion for detecting their mutual entanglement, which formally resembles the CHSH inequality~\cite{chsh}. 
However, contrary to the standard textbook case, the operations performed on each qubit are dynamically generated by the noise, with the strength of coupling to the environment
independently tailored for each qubit by an application of an appropriately chosen sequences of the control field pulses~\cite{Degen_RMP17,Szankowski_JPCM17}. 
With the ability to choose distinct settings for pulse sequences controlling each qubit, the local measurements performed after a period of noise-driven evolution result in four values of {\it correlators}, 
that are used to construct the CHSH-like inequality. 
As in the standard case, the inequality constructed in this way is a genuine criterion for entanglement as it is violated only by non-separable initial states of two qubits.
We also demonstrate that from the degree of the violation of such CHSH-like inequality one can deduce if the noise was Gaussian or not, answering thus a nontrivial question 
\cite{Norris_PRL16,Szankowski_JPCM17} about the statistics of environmental fluctuations. Finally, a remark is in order on the hierarchy of quantum correlations and the related non-classical effects
~\cite{cavalcanti2009experimental}.
The entanglement is a broad class of correlations, useful for quantum tasks such as the sub shot-noise metrology~\cite{giovannetti2004quantum,pezze2009entanglement}. 
Among the entangled states are those, where the Einstein-Podolsky-Rosen
steering is observed~\cite{epr,cavalcanti2009experimental}. An even more narrow subset is formed by states possesing the Bell correlations, which are responsible 
for the non-locality of quantum mechanics~\cite{bell,bell_rmp,Brunner_RMP14}.
Having this hierarchy in mind, we stress that the novel method we consider here, although bears a formal resemblance to the Bell test, 
cannot be used as a probe of non-locality of quantum mechanics, as the operations acting on each qubit are generated from a common source of noise.

This article is organized as follows. In Sec.~\ref{sec.chsh} we recall the main aspects of the standard CHSH inequality. Then, in Sec.~\ref{sec.noise} we introduce the decoherence-activated 
separability criterion. And so, in Sec.~\ref{sec.noise.observ} we show that the noise acting on the two-qubit systems can be used as a generator of local operations. To independently
control the strength of the local couplings, in Sec.~\ref{sec.noise.pulse}, we introduce the pulse sequence control method. In Sec.~\ref{sec.noise.chsh} we show the inequality is a criterion
for the entanglement of the initial state of two qubits, while in Sec.~\ref{sec.noise.sens} we discuss the family of states which violate this inequality, and thus are detected by the criterion. We conclude this part in Sec.~\ref{sec.noise.example} with an example of possible experimental implementation of this criterion. In Sec.~\ref{sec.crit} we show that the decoherence-activated 
separability criterion can be used to distinguish between the Gaussian and non-Gaussian noise. The conclusions are contained in Sec.~\ref{sec.conc}, while the Appendix presents details of some of the analytical calculations.

\section{The CHSH test as a separability criterion}\label{sec.chsh}

First, we briefly review the standard scheme of the CHSH test.

\subsection{The formulation of standard CHSH test}
Consider a pair of qubits, $A$ and $B$, initialized in a state described by the density matrix $\hat\varrho$, operating in the total Hilbert space of both qubits. The quantum-mechanical {\it correlator}
\begin{align}\label{eq:cor}
  E(\theta_A,\theta_B)&=\mathrm{Tr}\left[\hat{\boldsymbol{\sigma}}^{(A)}(\theta_A)
  	\otimes\hat{\boldsymbol{\sigma}}^{(B)}(\theta_B)\hat\varrho\,\right]
\end{align}
represents the average result of the measurement performed on the two qubit state $\hat\varrho$, with angles $\theta_A$ and $\theta_B$ parameterizing the choice of local observables
\begin{equation}
\hat{\boldsymbol{\sigma}}^{(Q)}(\theta) = \hat\sigma_x^{(Q)}\cos\theta+\hat\sigma_y^{(Q)}\sin\theta,
\end{equation}
where $Q=A,B$, and $\hat\sigma^{(Q)}_i$ ($i=x,y,z$) are the Pauli operators acting in the Hilbert subspace of qubit $Q$. This correlator can be used to construct the CHSH expectation value, parametrized by the choice of observable settings $\mathcal{S}=\{\alpha,\alpha',\beta,\beta'\}$
\begin{align}
\label{eq:CHSH_e_val}
\mathscr B_\mathcal{S}(\hat\varrho)\equiv E(\alpha,\beta)+E(\alpha,\beta')+E(\alpha',\beta)-E(\alpha',\beta'),
\end{align}
and the CHSH inequality
\begin{equation}
\label{eq:CHSH}
-2\leqslant \mathscr{B}_\mathcal{S}(\hat\varrho)\leqslant 2\,.
\end{equation}
When certain nontrivial conditions, such as space-like separation of measurement events and independence of measurement settings chosen randomly for the two qubits, are fulfilled, the violation of this inequality signifies the non-locality of state $\hat\varrho$, i.e., that the measured correlations described by the combination of correlators composing \eqref{eq:CHSH_e_val} cannot be explained by any local hidden variable models \cite{Brunner_RMP14}. However, we focus here on a much less demanding application of CHSH inequality to a simpler task: certifying if the state $\hat\varrho$ is entangled.

\subsection{CHSH test as a criterion for separability of two qubit state}
The inequality \eqref{eq:CHSH} can never be violated when the measurements are performed on a {\it separable} two-qubit state of the form 
\begin{equation}
\hat\varrho_{\mathrm{sep}} = \sum_k p_k\,\ketbra{\phi_k^{(A)}}{\phi_k^{(A)}}\otimes\ketbra{\phi_k^{(B)}}{\phi_k^{(B)}},
\end{equation}
where $\ket{\phi_k^{(Q)}}$ are---in general non-orthogonal---states of qubit $Q$, while
non-negative $p_k$'s add to unity \cite{Werner_PRA89,Horodecki_RMP09,Plenio_QIC07}.

On the other hand, some entangled two-qubit states do violate the CHSH inequality for a proper choice of settings $\mathcal{S}$. 
For instance, in the case of maximally entangled Bell state $\hat\varrho = |\Phi_+\rangle\langle \Phi_+|$, where
\begin{align}\label{eq:GHZ}
  \ket{\Phi_{\pm}}= \frac{\ket{{+}z^{(A)}}\ket{{+}z^{(B)}}\pm\ket{{-}z^{(A)}}\ket{{-}z^{(B)}}}{\sqrt 2}
\end{align}
(here $\ket{{\pm}z^{(Q)}}$ are the eigenstates of $\hat\sigma_z^{(Q)}$), the CHSH expectation value reaches the absolute maximum of $2\sqrt{2}\approx 2.83>2$ when $\alpha'=\alpha-\pi/2$, $\beta=\alpha+\pi/4$, $\beta'=\alpha+3\pi/4$, and arbitrary $\alpha$.

Thus, aside from all other contexts, like the Bell non-locality test, the CHSH inequality is a criterion for detecting the entanglement between two qubits. 
Indeed, as it was observed in e.g., \cite{Terhal_PLA00,Hyllus_PRA05,Schmied_Science16,wengerowsky2018field}, it is possible to repurpose the CHSH scheme as an {\it entanglement witness}. 
This is done by disassociating the correlator \eqref{eq:cor} from the initial state $\hat\varrho$ by defining the corresponding hermitian operators
\begin{equation}
\label{eq:corr_op}
\hat E(\theta_A,\theta_B) \equiv \hat{\boldsymbol{\sigma}}^{(A)}(\theta_A)\otimes\hat{\boldsymbol{\sigma}}^{(B)}(\theta_B),
\end{equation}
which are then combined into the {\it CHSH operator}
\begin{equation}
\label{eq:CHSH_op}
\hat{\mathscr{B}}_\mathcal{S} \equiv \hat E(\alpha,\beta)+\hat E(\alpha,\beta')+\hat E(\alpha',\beta)-\hat E(\alpha',\beta')\,.
\end{equation}
Then, one say that $\hat{\mathscr{B}}_\mathcal{S}$ has detected the entanglement in two-qubit state $\hat\varrho$, if its expectation value on that state exceeds the threshold for separable states $\max_{\hat\varrho_{\mathrm{sep}}}{|\mathrm{Tr}\,\hat{\mathscr{B}}_\mathcal{S}\hat\varrho_\mathrm{sep}|}=2$. In other words,
\begin{equation}
\label{eq:sep_crit}
\left(\text{If }|\mathrm{Tr}\,\hat{\mathscr{B}}_\mathcal{S}\hat\varrho| = |\mathscr{B}_\mathcal{S}(\hat\varrho)| > 2
	\,\text{, then $\hat\varrho$ is entangled.}\right)
\end{equation}
Formally, the CHSH operator $\hat{\mathscr{B}}_\mathcal{S}$ is not precisely an entanglement witness. Traditionally, the witness $\hat{\mathscr{W}}$ is defined as a hermitian operator such that its expectation value on any separable state is non-negative, i.e., $\mathrm{Tr}\,\hat{\mathscr{W}}\hat\varrho_\mathrm{sep}\geqslant 0$. Hence, the entanglement of state $\hat\varrho$ is witnessed by $\hat{\mathscr{W}}$ when $\mathrm{Tr}\,\hat{\mathscr{W}}\hat\varrho <0$. Of course, it is a trivial matter to construct a proper witness out of the CHSH operator, by simply combining $\hat{\mathscr{B}}_\mathcal{S}$ with operator proportional to the identity $\hat{\mathds{1}}$. In order to cover all cases when the criterion \eqref{eq:sep_crit} tests positive, one should define two classes of witnesses: $\hat{\mathscr{W}}_\mathcal{S}\equiv 2\hat{\mathds{1}}-\hat{\mathscr{B}}_\mathcal{S}$, so that $\mathrm{Tr}\,\hat{\mathscr{W}}_\mathcal{S}\hat\varrho < 0$ is equivalent to $\mathscr{B}_\mathcal{S}(\hat\varrho)>2$, and $\hat{\mathscr{W}}_\mathcal{S}'\equiv 2\hat{\mathds{1}}+\hat{\mathscr{B}}_\mathcal{S}$, for which $\mathrm{Tr}\,\hat{\mathscr{W}}_\mathcal{S}'\hat\varrho<0$ corresponds to $\mathscr{B}_\mathcal{S}(\hat\varrho)<-2$. However, the latter class of witnesses is actually superfluous, because it can be transformed into the former class by a proper choice of the settings, namely $\hat{\mathscr{W}}_{\{\alpha+\pi/2,\alpha'+\pi/2,\beta+\pi/2,\beta'+\pi/2\}} = \hat{\mathscr{W}}_\mathcal{\{\alpha,\alpha',\beta,\beta'\}}'$.

\section{Decoherence-activated separability criterion}\label{sec.noise}
Below, the decoherence-activated separability criterion is derived from dynamically driven CHSH-like scheme. The key difference between standard CHSH test and this scheme, 
boils down to one essential modification: the choice of observable settings is taken over by the noise coupled to qubits, that were initialized in a state to be discriminated by the criterion.  The source
of noise is common for both qubits, thus what we propose here is the entanglement test which cannot be interpreted as a probe of non-locality.

\subsection{The CHSH test with the noise induced choice of local observables}\label{sec.noise.observ}
We now consider a scheme where the dynamics induced by the environmental noise are incorporated into the setup that formally resembles the traditional CHSH separability criterion. The two qubits, positioned at $\x_A$ and $\x_B$, initially prepared in the state $\hat\varrho$, are subjected to the external noise represented by a stochastic vector field $\boldsymbol{\xi}(\mathbf{r}, t)$. For convenience the noise is assumed to be stationary and have zero average. More importantly, it also assumed that the noise is weak as compared to the unperturbed energy splitting of each qubit, described by their free Hamiltonians $\hat H_0 =\sum_{Q=A,B}\Omega_Q \hat\sigma_z^{(Q)}/2$, possibly with a distinct physical orientation of the $z$ axes for each qubit. Under this assumption, and additionally when noise has little spectral content at high frequencies $\approx \! \Omega_{Q}$, the noise-induced transitions between eigenstates of $\hat\sigma_z^{(Q)}$ occur on much longer timescale than dephasing of superpositions of these eigenstates. Consequently, we ignore from now on
the influence of components of  $\boldsymbol{\xi}(\mathbf{r}_{A/B}, t)$  transverse to the quantization axis of qubits $A/B$, and focus only on the fluctuations of longitudinal components $\xi_{z}(\mathbf{r}_{A/B})$ affecting the coherence of the qubits. The influence of transverse components can be approximately taken into account as a second-order contribution to longitudinal fluctuations $\sim \xi_{\perp}^2/2\Omega_{Q}$ that can be incorporated into effective $\xi_{z}$ (note that even for $\xi$ being a Gaussian process, $\xi^2$ is not Gaussian \cite{Makhlin_PRL04,Cywinski_PRA14}, which can have interesting consequences for results discussed below, see especially Sec.~\ref{sec.crit}). After suppressing the $z$ subscript of $\xi_{z}(\mathbf{r}_{A/B},t)$ for simplicity, the qubit-noise coupling is then given by
\begin{align}\label{eq:int_ham}
  \hat V=\frac{1}{2}\,\xi(\x_A, t)\,\hat\sigma_z^{(A)}+\frac{1}{2}\,\xi(\x_B, t)\,\hat\sigma_z^{(B)},
\end{align}
which commutes with the free Hamiltonian $[\hat H_0,\hat V]=0$. The qubits are allowed to evolve for duration $T$, and then the measurement is performed with both local observables fixed to $\hat{\boldsymbol{\sigma}}^{(Q)}(0)=\hat\sigma_x^{(Q)}$.

For each realization of $\xi$ drawn from the probability distribution functional $P(\xi)$, the qubits undergo unitary evolution
\begin{align}\label{eq:loc_trans}
  \hat U_\xi=\exp\left(-i \frac{\alpha_\xi}{2} \hat \sigma_z^{(A)}\right)\otimes \exp\left(-i \frac{\beta_\xi}{2} \hat \sigma_z^{(B)}\right), 
\end{align}
where the phases accumulated on each qubit are given by
\begin{align}
\nonumber
  \alpha_\xi &= \int_0^T \left(\Omega_A+\xi(\mathbf r_A,t)\right)dt\,,\\
\label{eq:angles}
  \beta_\xi &= \int_0^T \left(\Omega_B+\xi(\mathbf r_B,t)\right)dt\,.
\end{align}
Accordingly, the quantum-mechanical expectation value of the measurement result is given by
\begin{align}
\nonumber
&\mathrm{Tr}\left[ \hat\sigma_x^{(A)}\otimes\hat\sigma_x^{(B)}\hat U_\xi\hat\varrho\hat U^\dagger_\xi\right]=
\mathrm{Tr}\left[\left(\hat U^\dagger_\xi \hat\sigma_x^{(A)}\otimes\hat\sigma_x^{(B)}\hat U_\xi\right)\hat\varrho\right]\\
\label{eq:corr_per_trajectory}
&=\mathrm{Tr}\left[\hat{\boldsymbol{\sigma}}^{(A)}(\alpha_\xi)\otimes\hat{\boldsymbol{\sigma}}^{(B)}(\beta_\xi)\hat\varrho\right]
	=\mathrm{Tr}\,\hat E(\alpha_\xi ,\beta_\xi )\hat\varrho\,,
\end{align}
and by comparing the above formula with Eq.~\eqref{eq:corr_op} we see that, when examined per trajectory of the noise, the Heisenberg picture of the transformation (\ref{eq:loc_trans}) effectively takes over the choice of the angles of local observables, and sets them randomly to $\alpha_\xi$ and $\beta_\xi$ \cite{Szankowski_QIP15}. 

In order to obtain the actual average measurement result fit to be represented by a correlator, Eq.~\eqref{eq:corr_per_trajectory} must be averaged over realizations of $\xi$, giving
\begin{equation}\label{eq:bad_av_corr}
  \overline E= \mathrm{Tr}\,\int\mathcal{D}\xi P(\xi)\hat E(\alpha_\xi,\beta_\xi)\hat\varrho \equiv \mathrm{Tr}\,\hat{\overline{E}}\hat\varrho.
\end{equation}
Note that neither $\overline{E}$, nor its operator counterpart $\hat{\overline{E}}$, are useful correlators for the purpose of construction of Bell test or separability criterion, as they do not allow for manipulation of local settings. We enable this crucial element by introducing into our scheme the qubit control based on dynamical-decoupling techniques \cite{Szankowski_JPCM17}.

\subsection{Adjusting the local settings with the qubit control}\label{sec.noise.pulse}

\begin{figure}[t]
\centering
\includegraphics[width=\columnwidth]{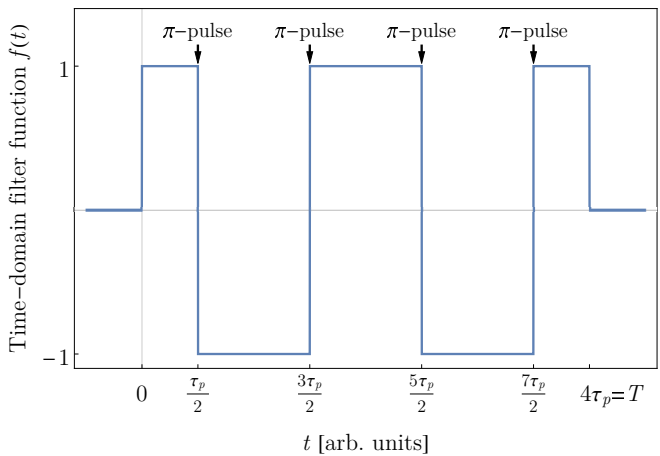}
\caption{The time-domain filter function $f(t)$. The example depicted in the figure was created by an $n=4$ pulse Carr-Purcell sequence, defined by the interpulse delay $\tau_p$, and the pulse timings: $t_1 = \tau_p/2$, $t_2=t_1+\tau_p$, $\ldots$, $t_k = t_{k-1}+\tau_p$, $\ldots$, $t_{n}=t_{n-1}+\tau_p/2$. The total duration of the sequence (and consequently, of the filter function) is $T=n\tau_p$.}
\label{fig:time-dom_filter_examples}
\end{figure}
During the evolution, each qubit is now individually subjected to a sequence of pulses of external field, that cause effectively instantaneous $\pi$-rotations (i.e., spin-flips). Consequently, the angles accumulated over a course of single realization of $\xi$ are modified according to
\begin{align}
\nonumber
&\alpha_{\xi}(a) = \int_0^T  f_{a}^{(A)}(t)\,\xi(\mathbf r_A,t)\, dt,\\
\label{eq:angles_filtered_xi}
&\beta_{\xi}(b) = \int_0^T f_b^{(B)}(t)\, \xi(\mathbf r_B,t)\, dt,
\end{align}
where $f_{a/b}^{(A/B)}(t)$ are the time-domain filter functions, encapsulating the effects of pulse sequences applied to each qubit \cite{deSousa_TAP09,Cywinski_PRB08,Szankowski_JPCM17}.
The filter functions have a form of square waves, switching between $\pm 1$ at moments when a pulse causes the spin-flip, see Fig.~\ref{fig:time-dom_filter_examples}. 

Here we choose to control our qubits with Carr-Purcell sequences \cite{Szankowski_JPCM17}, that are defined by the interpulse delay $\tau_p$ and the series of pulse timings: $t_1=\tau_p/2, t_2 =t_1+\tau_p,\ldots t_k = t_{k-1}+\tau_p,\ldots,t_n = t_{n-1}+\tau_p/2 $, and $T= t_n+\tau_p/2=n\tau_p$. The filter function induced by such a sequence is periodic, and it is shaped as a square wave that oscillates with well defined frequency $\omega_p = \pi/\tau_p$. Therefore, the filter lets through the harmonic components of the noise that oscillate with frequencies around $\omega_p$ (and its odd multiples, because filter function is a square wave, not pure harmonic wave), and suppresses all the other frequencies. The width of the passband of the filter is of the order of $T^{-1}$, hence the filter is more precise for long evolution durations (equivalently, larger number of pulses). For example, in transition form Eq.~\eqref{eq:angles} to \eqref{eq:angles_filtered_xi}, the contribution from time-independent free qubit Hamiltonian (the splittings $\Omega_A$ and $\Omega_B$) has been eliminated, because the passband of Carr-Purcell sequences with finite interpulse delay cannot be centered around $\omega_p=0$.

For our purposes it is enough to consider two distinct pulse sequences per party, and an additional one only for party $B$; each of those sequences is playing a role of measurement setting in the standard CHSH scheme. The first setting, labeled as ``On'', is realized by choosing such $\omega_p$ that the passband of the filter aims at harmonic of the noise that is characterized by significant average intensity. To be more concrete, a convenient measure of the spectral contents of the zero average, stationary noise is its {\it spectral density} (or simply {\it spectrum}) \cite{Kogan,Szankowski_JPCM17}, defined as $S_{Q}(\omega) = \int_{-\infty}^\infty dt e^{-i\omega t}\int\mathcal{D}\xi P(\xi) \xi(\mathbf{r}_Q, t)\xi(\mathbf{r}_Q, 0)$; it describes the distribution of power into frequency components composing the noise at the location of the respective qubit. The reconstruction of this function is the main objective of the dynamical-decoupling-based noise spectroscopy \cite{Szankowski_JPCM17,Degen_RMP17}. In its most straightforward implementation, it is facilitated by the feature of the pure dephasing under the action of pulse sequences, where the decoherence rate is proportional to the spectral density evaluated at filter frequency $\omega_p$. By measuring this rate for a spread of settings of $\omega_p$ one can recover these values and ultimately recreate the course of $S_Q(\omega)$. Here we assume that the spectral density has already been characterized and we intend to utilize the pulse sequences to control the rate of dephasing by aiming the filter frequencies at different parts of noise spectrum. Hence, the setting ``On'' would correspond to the choice of $\omega_p$ for the respective pulse sequence, so that $S_{A/B}(\omega_p)$ has an appreciable value. In contrast, the second setting labeled as ``Off'', corresponds to such a choice of $\omega_p$ that $S_{A/B}(\omega_p)\approx 0$. In other words, with setting ``Off'' the pulse sequence decouples the qubit form the noise, while the setting ``On'' is designed to produce the opposite effect: to couple the qubit to an intense portion of the noise. 

Let us note here, that in principle the ``On'' setting could also be realized without exerting additional control over qubit, i.e., by allowing the qubit to evolve freely due to the action of unfiltered noise. The drawback of such a simplistic approach is the lack of fine control over the rate of the noise-induced dephasing.

With the first two control sequences included, the ability to manipulate local settings has been restored, and the correlator now reads
\begin{widetext}
\begin{align}
\label{eq:av_corr_form}
\overline E(a,b) &=\mathrm{Tr}\,\int\mathcal{D}\xi P(\xi)\hat{E}\Big(\alpha_\xi(a),\beta_\xi(b)\Big)\hat\varrho\equiv \mathrm{Tr}\,\hat{\overline{E}}(a,b)\hat\varrho\,,\\[.3cm]
\nonumber
  \hat{\overline E}(a,b)&= \exp[\, {-}\chi_A(a)-\chi_B(b)-2\chi_{[AB]}(a,b)\, ]\,\frac{\hat E(0,0)+\hat E(\tfrac{\pi}{2},\tfrac{\pi}{2})}{2}\\
\label{eq:av_corr_op}
	&+\exp[\, -\chi_A(a)-\chi_B(b)-2\chi_{\{AB\}}(a,b)\, ]\,\frac{\hat E(0,0)-\hat E(\tfrac{\pi}{2},\tfrac{\pi}{2})}{2}
\end{align}
\end{widetext}
with $a$ and $b$ equal to ``On'' or ``Off'' and $\hat E(\theta_A,\theta_B)$ given by Eq.~\eqref{eq:corr_op}. The arguments of the exponential functions resulting from the averaging over noise realizations, has been split into distinct parts. First, we have {\it local attenuation functions} $\chi_{A}(a)$ and $\chi_{B}(b)$, that depend only on the setting of sequence applied locally to qubit $A$ and $B$ respectively, and it has a form of infinite series of auto-correlation functions of $\alpha_\xi(a)$ and $\beta_\xi(b)$ of all orders (i.e., the cumulant series of the respective stochastic phase) \cite{Szankowski_JPCM17}. For example, the second order auto-correlation function is given by the inverse Fourier transform of  the spectral density we discussed previously. Then, the {\it non-local} attenuation functions, $\chi_{\{AB\}}(a,b)$ and $\chi_{[AB]}(a,b)$, are formed by an analogous series, but composed of cross-correlation functions between $\alpha_\xi(a)$ and $\beta_\xi(b)$, in the case of the former, and between $\alpha_\xi(a)$ and $-\beta_\xi(b)$, in the case of the latter. Consequently, the non-local attenuation functions depend on both settings. Appendix \ref{app:avg_corr} contains the derivation of this result, and it must be stressed that it was obtained under an additional assumption that the noise, on average, does not cause the phase to shift; in technical terms, this condition is equivalent to the assumption that all attenuation functions are purely real. This holds for noise with Gaussian statistics, and for all non-Gaussian noises that have vanishing odd-order cumulants \cite{Szankowski_JPCM17}. The latter class includes Random Telegraph Noise \cite{Galperin_03,Szankowski2013,Ramon_PRB2015} that often affects solid-state based qubits, and which is discussed in Sec.~\ref{sec.crit}. 

Finally, the third option for the setting $b$ mentioned above, labeled as ``$\onstar{}$'', is a variation on ``$\on{}$'': it is realized by a control sequence with the same $\omega_p$ as for ``$\on{}$'', but additionally appended with a single $\pi$ pulse at the very beginning of the evolution. Consequently, the time-domain filter function produced by this pulse sequence is given by $f_{\onstar{}}^{(B)}(t) = -f_{\on{}}^{(B)}(t)$, which leads to $\beta_\xi(\onstar{})=-\beta_\xi(\on{})$. Therefore, when the setting $b$ for qubit $B$ is chosen to be ``$\onstar{}$'' instead of ``$\on{}$'', the non-local attenuation functions get transformed according to the following rules
\begin{subequations}
\begin{align}
\chi_{\{AB\}}(a,\onstar{}) &= \chi_{[AB]}(a,\on{})\,,\\
\chi_{[AB]}(a,\onstar{}) &= \chi_{\{AB\}}(a,\on{})\,,
\end{align}
\end{subequations}
while the local attenuation function remains unchanged, $\chi_B(\onstar{})=\chi_B(\on{})$. This ability to transmute non-local attenuation functions one into another is the only purpose for introducing this setting. The usefulness of such a tool will become apparent, when we proceed with the construction of entanglement witnesses and separability criteria out of correlators \eqref{eq:av_corr_op}.

\subsection{Separability criteria}\label{sec.noise.chsh}
The exact value of attenuation functions depends on the probability distribution $P(\xi)$, and in general is it impossible to express it in a closed form, with a notable exception of Gaussian noise 
\cite{Cywinski_PRB08,Szankowski_JPCM17}. However, this difficulty posses no real hindrance for our proceedings. Assuming that the setting ``Off'' realizes an efficient dynamical decoupling, the correlation functions of the noise local to the decoupled qubit can be set to zero. Then, the attenuation function local to this qubit vanishes, which in turn implies that the non-local attenuation functions disappear as well. Thus, the possible options for local settings generate the following values of the correlators (from this point we shall omit the arguments of the non-zero attenuation functions for clarity):
\begin{subequations}
\label{eq:ON_OFF_correlators}
\begin{align}
\hat{\overline{E}}(\text{Off},\text{Off}) &\approx  \hat E(0,0)\,,\\[.2cm]
\hat{\overline{E}}(\text{On},\text{Off}) &\approx e^{-\chi_A} \hat E(0,0)\,,\\[.2cm]
\hat{\overline{E}}(\text{Off},\text{On}) &=\hat{\overline{E}}(\off{},\onstar{}) \approx e^{-\chi_B}\hat E(0,0)\\[.2cm]
\nonumber
\hat{\overline{E}}(\on{},\on{})&= e^{-\chi_A-\chi_B-2\chi_{[AB]}}\frac{\hat E(0,0)+\hat E(\tfrac{\pi}{2},\tfrac{\pi}{2})}{2}\\
\label{subeq:ON_OFF_corr:On_On}
	&+e^{-\chi_A-\chi_B-2\chi_{\{AB\}}}\frac{\hat E(0,0)-\hat E(\tfrac{\pi}{2},\tfrac{\pi}{2})}{2},\\[.2cm]
\nonumber
\hat{\overline{E}}(\on{},\onstar{}) &=e^{-\chi_A-\chi_B-2\chi_{\{AB\}}}\frac{\hat E(0,0)+\hat E(\tfrac{\pi}{2},\tfrac{\pi}{2})}{2}\\
\label{subeq:ON_OFF_corr:On_oN}
	&+e^{-\chi_A-\chi_B-2\chi_{[AB]}}\frac{\hat E(0,0)-\hat E(\tfrac{\pi}{2},\tfrac{\pi}{2})}{2},
\end{align}
\end{subequations}
where the only difference between \eqref{subeq:ON_OFF_corr:On_On} and \eqref{subeq:ON_OFF_corr:On_oN} is the sign change in the terms proportional to $\hat E(\tfrac{\pi}{2},\tfrac{\pi}{2})$.

The correlators \eqref{eq:ON_OFF_correlators} are now combined into two hermitian {\it \CHSHlike{} operators}, that are analogous to the class of CHSH operators $\hat{\mathscr{B}}_\mathcal{S}$:
\begin{subequations}
\label{eq:avg_CHSH_op}
\begin{align}
  \nonumber
  \hat{\overline{\mathscr{B}}}_\Phi \equiv&\ \hat{\overline{E}}(\off{},\off{})+\hat{\overline{E}}(\on{},\off{})\\
  \label{subeq:avg_CHSH_op:GHZ_witness}
  &\ +\hat{\overline{E}}(\off{},\on{})-\hat{\overline{E}}(\on{},\on{})\,,\\[.2cm]
  \nonumber
  \hat{\overline{\mathscr{B}}}_\Psi \equiv&\ \hat{\overline{E}}(\off{},\off{})+\hat{\overline{E}}(\on{},\off{})\\
  \label{subeq:avg_CHSH_op:singlet_witness}
  &\ +\hat{\overline{E}}(\off{},\onstar{})-\hat{\overline{E}}(\on{},\onstar{})\,.
\end{align}
\end{subequations}
Two fundamental properties, relevant to potential application as a separability criterion, can be inferred from the form of these operators (see Appendix \ref{app:prop_avg_CHSH_op}): (i) the expectation value of $\hat{\overline{\mathscr{B}}}_\mathcal{S}$ (with $\mathcal{S}=\Phi,\Psi$) on any separable state is bounded by the threshold of the standard CHSH expectation value, $|\mathrm{Tr}\,\hat{\overline{\mathscr{B}}}_\mathcal{S}\hat\varrho_\mathrm{sep}|\leqslant 2$, which is smaller then the overall maximum of $2\sqrt{2}$, and (ii) for $\chi_{\{AB\}}=\chi_{[AB]}=0$, i.e., in the case when noise field values at the location of each qubit are completely uncorrelated, the expectation value of $\hat{\overline{\mathscr{B}}}_\mathcal{S}$ on {\it any} state $\hat\varrho$ never exceeds the threshold for separable states.

Therefore, as long as the noises affecting each qubit are correlated, the expectation value of operators \eqref{eq:avg_CHSH_op} can serve as a proper separability criteria
\begin{equation}\label{eq:avg_sep_crit}
\left(\text{If }|\mathrm{Tr}\,\hat{\overline{\mathscr{B}}}_{\Phi/\Psi}\hat\varrho|>2\,\text{, then $\hat\varrho$ is entangled.}\right)
\end{equation}

\begin{figure}[t]
	\centering
	\includegraphics[width=\columnwidth]{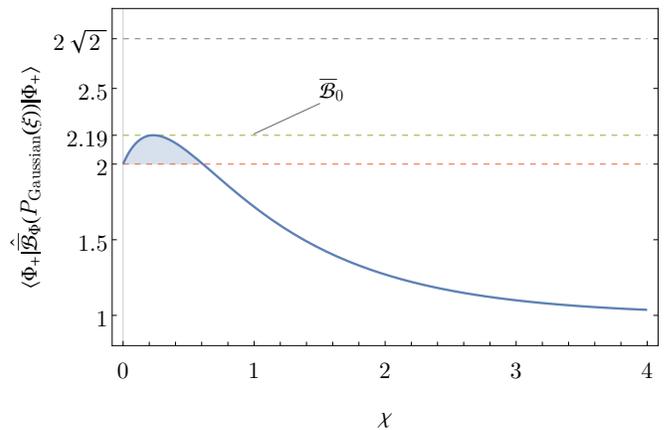}
	\caption{The performance of the entanglement criterion \eqref{eq:avg_sep_crit} testing the Bell state $|\Phi_+\rangle$, depending on the attenuation function $\chi$ induced by the perfectly correlated Gaussian noise, as given by Eq.~\eqref{eq:avg_CHSH_Gauss}. The criterion is entanglement positive if the curve passes over the threshold 
	$\max_{\hat\varrho_\mathrm{sep}}|\mathrm{Tr}\,\hat{\overline{\mathscr{B}}}_\Phi \hat\varrho_\mathrm{sep}|=2$ (red horizontal dashed line). The range of $\chi$ for which the entanglement is detected is indicated by the undercurve shading. The maximum value of the criterion $\overline{\mathscr{B}}_0\approx 2.19$ (see Eq.~\eqref{eq:bound}) is achieved for $\chi=\ln(2)/3$; it is significantly smaller than the overall maximum of $2\sqrt{2}$.}
	\label{fig:gauss-crit}
\end{figure}

Of course, although the criterion never yields false positives, it would be useless unless it is also capable of producing actual true positives. The detection of entanglement with criterion \eqref{eq:avg_sep_crit} can be demonstrated in the most transparent manner for the case of the system initialized in $|\Phi_{\pm}\rangle$ state and perfectly correlated Gaussian noises driving the evolution. When the noises are Gaussian the series constituting the attenuation functions are truncated at the second order correlation functions: 
\begin{subequations}
\label{eq:gauss_att_func}
\begin{align}
\chi_A(a)&=\frac{1}{2}\int\mathcal{D}\xi P_\mathrm{Gaussian}(\xi) \alpha_\xi^2(a),\\[.3cm]
\chi_B(b)&=\frac{1}{2}\int\mathcal{D}\xi P_\mathrm{Gaussian}(\xi)\beta_\xi^2(b),\\[.3cm]
\nonumber
\chi_{AB}(a,b) &= \frac{1}{2}\int\mathcal{D}\xi P_\mathrm{Gaussian}(\xi) \alpha_\xi(a)\beta_\xi(b)\\
&=\chi_{\{AB\}}(a,b)= -\chi_{[AB]}(a,b).
\end{align}
\end{subequations}
The perfect correlation means that the noises affecting each qubit are exactly the same for each realization of stochastic process, i.e., 
$\xi(\mathbf{r}_A, t)= \xi(\mathbf{r}_B,t)$ (the opposite of uncorrelated noises). 
In addition, if pulse sequences applied to each qubit are also the same, then the cross-correlation equals the auto-correlations, 
and consequently the non-local and local attenuation functions become identical, $\chi_A{(\text{On})}=\chi_B{(\text{On})} = \chi_{AB}(\text{On},\text{On})\equiv \chi$. 
In these circumstances, the expectation value of $\Phi$-operator \eqref{subeq:avg_CHSH_op:GHZ_witness} reads
\begin{equation}
\label{eq:avg_CHSH_Gauss}
 |\langle\Phi_{\pm}| \hat{\overline{\mathscr{B}}}_\Phi\big(P_\mathrm{Gaussian}(\xi)\big)|\Phi_{\pm}\rangle|=1+2e^{-\chi}-e^{-4\chi}.
\end{equation}
This reaches the maximal value of
\begin{equation}\label{eq:bound}
  \overline{\mathscr{B}}_{0} = 1 + 2^{2/3}-2^{-4/3} = 1+3\times 2^{-4/3} \approx 2.19 > 2, 
\end{equation}
for $\chi= \ln(2)/3$, and simultaneously is a positive result for detection of entanglement in states $|\Phi_\pm\rangle$ (see Fig.~\ref{fig:gauss-crit}).  
On the other hand, the expectation value of $\Psi$-operator \eqref{subeq:avg_CHSH_op:singlet_witness} on the same state gives only
\begin{equation}
|\langle\Phi_\pm|\hat{\overline{\mathscr{B}}}_\Psi\big(P_\mathrm{Gaussian}(\xi)\big)|\Phi_\pm\rangle| = 2e^{-\chi}\leqslant 2\,,
\end{equation}
as it fails to reveal the presence of entanglement. However, for the other two Bell states,
\begin{equation}
|\Psi_\pm\rangle = \frac{\ket{{+}z^{(A)}}\ket{{-}z^{(B)}}\pm\ket{{-}z^{(A)}}\ket{{+}z^{(B)}}}{\sqrt 2}\,,
\end{equation}
the capabilities of \CHSHlike{} operators are reversed: $|\langle\Psi_\pm|\hat{\overline{\mathscr{B}}}_\Psi\big(P_\mathrm{Gaussian}(\xi)\big)|\Psi_\pm\rangle|=1+2e^{-\chi}-e^{-4\chi}$, while $|\langle\Psi_\pm|\hat{\overline{\mathscr{B}}}_\Phi\big(P_\mathrm{Gaussian}(\xi)\big)|\Psi_\pm\rangle|=2e^{-\chi}$. This example justifies the need for introduction of ``$\onstar{}$'' setting and the two classes of \CHSHlike{} operators. It is not dissimilar to the issue with the standard CHSH separability criterion, where the capability to detect a given type of entanglement depended on the settings of local observables $\mathcal{S}=\{\alpha,\alpha',\beta,\beta'\}$.

Similarly to standard CHSH operators, the classes of \CHSHlikeAD{} operators can serve as a constituents of entanglement witnesses. In order to encompass all cases of positives identifiable by criterion \eqref{eq:avg_sep_crit}, one requires four classes of witnesses: $\hat{\mathscr{W}}_{\Phi_\pm}= 2\hat{\mathds{1}}\mp\hat{\overline{\mathscr{B}}}_\Phi$, that are capable of discriminating $|\Phi_\pm\rangle$, and $\hat{\mathscr{W}}_{\Psi_\pm}= 2\hat{\mathds{1}}\mp\hat{\overline{\mathscr{B}}}_\Psi$, tailored for witnessing the entanglement in $|\Psi_\pm\rangle$. Note that, unlike the case of standard CHSH scheme, none of these classes of witnesses are superfluous.

As a side note, it is possible to understand the role of noise correlations in the performance of this separability criteria in terms of quantum Fisher information and its physical interpretation as a measure of state's susceptibility to certain transformations, that sets the ``speed'' limits for its evolution \cite{PhysRevA.85.052127,PhysRevA.89.012307,Wasak_PhD_2015}. For example, in the case of perfectly correlated noise, evolution of the investigated two-qubit state is generated by a global angular momentum operator $\hat J_z = (\hat\sigma_z^{(A)}+\hat\sigma_z^{(B)})/2$. For unitary evolution generated by this operator it is known that when quantum Fisher information is greater than $2$, the two-qubit state has to be entangled. Hence, in the case of unitary evolution the classical susceptibility limit for the phase transformation is well established. In the case of evolution due to noise, it is not clear what is this limit, especially when the noises are only partially correlated. For pure dephasing discussed here, the relevant quantum Fisher information is proportional to the correlators $\overline{E}(a,b)$ \cite{Szankowski_PRA14}, and so, our separability criteria can be understood as a convenient way to compare the susceptibilities to decoherence in four characteristic situations with a single number. The key element is the comparison between the ``individual decoherences'' $\overline{E}(\mathrm{On},\mathrm{Off})$ and $\overline{E}(\mathrm{Off},\mathrm{On})$, versus the susceptibility to ``collective decoherence'' $\overline{E}(\mathrm{On},\mathrm{On})$ [the term $\overline{E}(\mathrm{Off},\mathrm{Off})$ plays the role of the reference level, which becomes non-trivial when the dynamical decoupling is not perfect and $\chi(\mathrm{Off},\mathrm{Off})>0$]. Such a comparison might reveal entanglement, because the susceptibility of classical states is simply a sum of susceptibilities of its constituents, while the  ``collective'' susceptibility in the presence of quantum correlations can be higher than that. Here is the point where the noise correlations come into play: in order to be able to induce the ``collective'' mode of decoherence, one needs some correlations between noises driving each qubit, otherwise even $\overline{E}(\mathrm{On},\mathrm{On})$ would measure only ``individual'' susceptibilities [e.g., for perfect dynamical decoupling one would get $\overline{E}(\mathrm{On},\mathrm{On}) \propto \overline{E}(\mathrm{On},\mathrm{Off})\overline{E}(\mathrm{Off},\mathrm{On})$, which is redundant with the information on susceptibility of ``individual'' decoherences].

It is interesting to note that a result identical to the one from Eq.~(\ref{eq:bound}) was obtained in \cite{banaszek3}, in which a maximal violation of  CHSH inequality for two-mode squeezed vacuum state produced in a process of nondegenerate optical parametric amplification was considered. The four ``measurement settings'' in that paper corresponded to four different manipulations, in a form of phase-space displacements, of the tested state followed by a measurement of a product of displaced parity operators. The mathematical equivalence of results follows from a formal analogy between calculation of expectation values of parity operator on Gaussian states of photon field and averaging of Eq.~\eqref{eq:av_corr_form} over the realizations of noise with Gaussian statistics.

\subsection{The sensitivity of \CHSHlike{} separability criterion}\label{sec.noise.sens}

In previous section it was demonstrated that the \CHSHlike{} criterion \eqref{eq:avg_sep_crit} is at least capable of distinguishing maximally entangled Bell states. The question is, how sensitive the criterion is, i.e., how large is the set of entangled states that would trigger a positive result. Instead of trying to identify the exact boundaries of such a set, we will gauge this sensitivity by testing the performance of the criterion on a family of Werner states
\begin{equation}
\hat\varrho_p = \frac{1}{4}(1-p)\hat{\mathds{1}}+p\,|\Psi_-\rangle\langle\Psi_-|\,,
\end{equation}
that are parametrized by $p\in [0,1]$.

According to Peres-Horodecki separability criterion \cite{Aolita_RPP15}, which is known to have $100\%$ sensitivity for two-qubit states (i.e., it is capable of detecting all entangled two-qubit states), Werner state $\hat\varrho_p$ is entangled for $p>p_0=1/3$. In comparison, the standard CHSH criterion \eqref{eq:sep_crit} for optimally chosen settings, detects entanglement if $p>p_\mathrm{CHSH}=1/\sqrt 2\approx 0.71$ \cite{Aolita_RPP15}. Therefore, CHSH criterion is not perfectly sensitive, as it is capable to positively identify only a fraction of entangled $\hat\varrho_p$ states,
\begin{equation}
\sigma_\mathrm{CHSH}=\frac{1-p_\mathrm{CHSH}}{1-p_0}100\%\approx 44\%\,.
\end{equation}

The \CHSHlike{} criterion (with the optimal setting $\mathcal S = \Psi$) yields a positive result when
\begin{equation}\label{eq:avg_crit_Werner}
 |\mathrm{Tr}\,\hat{\overline{\mathscr{B}}}_{\Psi}\hat\varrho_{p}| =p|\langle\Psi_-|\hat{\overline{\mathscr{B}}}_\Psi|\Psi_-\rangle|>2\,,
\end{equation}
which leads to the threshold value of $p$ for detecting entanglement in Werner states,
\begin{equation}
p>p_{\bar{\mathscr{B}}} = \frac{2}{|\langle\Psi_-|\hat{\overline{\mathscr{B}}}_\Psi\big(P(\xi),f^{(A)}_\on{},f^{(B)}_\on{}\big)|\Psi_-\rangle|}\,.
\end{equation}
The arguments of operator $\hat{\overline{\mathscr{B}}}_\Psi$ have been included here to underline that the threshold, and consequently the sensitivity of \CHSHlike{} criterion, depends on the statistics of the noise $P(\xi)$, as well as the choice of qubit control parameters in the ``$\on{}$'' settings. For example, in the case of perfectly correlated Gaussian noise, the threshold can be only as low as $p_{\bar{\mathscr{B}}}=p_\mathrm{Gaussian} = 2/\overline{\mathscr{B}}_0\approx 0.91$, which gives sensitivity of $\sigma_\mathrm{Gaussian}=\frac{1-p_\mathrm{Gaussian}}{1-p_0}100\% \approx 13.5\%$.

In addition, one can observe two universal (i.e., independent of the statistics of the noise) properties of the criterion: (i) in the regime of weak dephasing, when $|\chi_Q|\ll 1$---which in turn imply that $|\chi_{\{AB\}}|\ll 1$ and $|\chi_{[AB]}|\ll 1$---if the non-local attenuation function is {\it positive}, $0<\chi_{\{AB\}}\ll 1$, the threshold is always smaller than one
\begin{align}
p_{\bar{\mathscr{B}}}\approx p_\mathrm{weak}=(1+\chi_{\{AB\}})^{-1}  \,.
\end{align}
Therefore, independently of the noise statistics, the criterion is capable of detecting some entangled states, but only with a low sensitivity, $\sigma_\mathrm{weak} = (1-p_\mathrm{weak})/(1-p_0)\approx 3\chi_{\{AB\}}/2$. (ii) In the opposite regime of strong dephasing, when $\chi_Q\sim 1$, the sensitivity of the criterion drops to zero, because
\begin{align}
\nonumber
p|\langle\Psi_-|\hat{\overline{\mathscr{B}}}_\Psi|\Psi_-\rangle|
	\approx p(1+2e^{-1}-e^{-2}e^{-2\chi_{\{AB\}}})&\\
\leqslant p(1+2e^{-1})\simeq p\times 1.74<2&,
\end{align}
as at this point the erosion of quantum correlations due to decoherence has become more of an inhibitor instead a catalyst.

\subsection{An example of experimental implementation}\label{sec.noise.example}

\begin{figure}[t]
\centering
\includegraphics[width=\columnwidth]{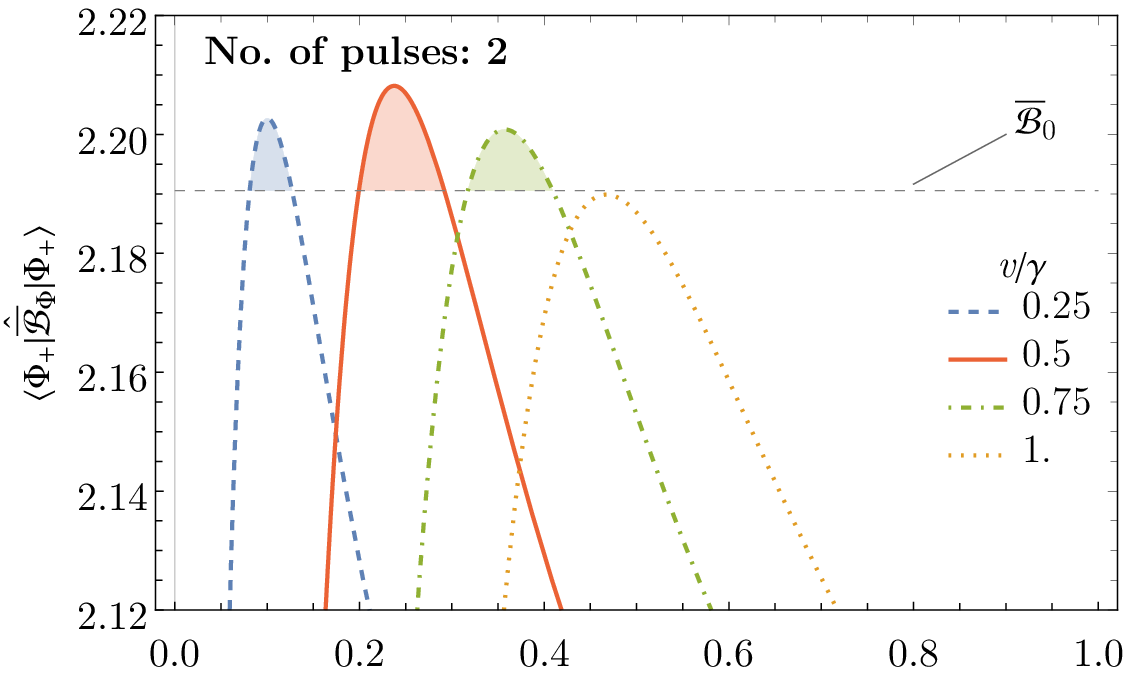}
\includegraphics[width=\columnwidth]{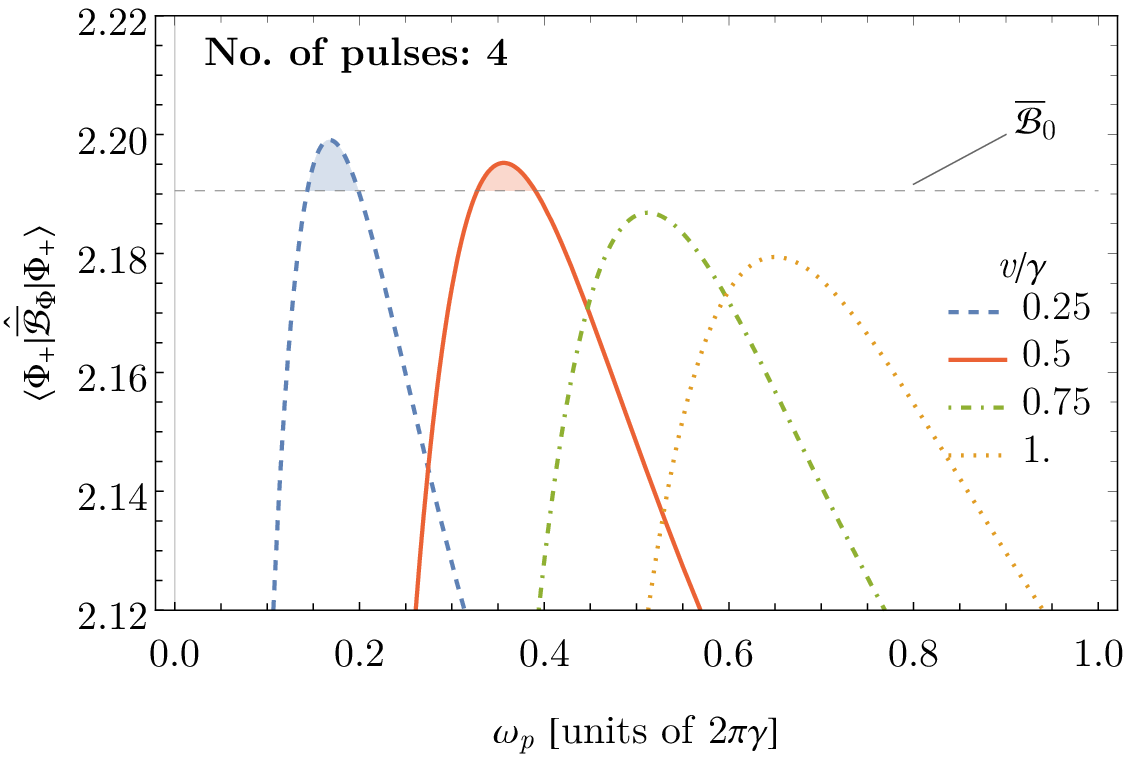}
\caption{Examples of application of non-Gaussianity criterion \eqref{eq:non-gauss_crit} to the case of perfectly correlated random telegraph noises (RTNs), for various configurations of noise and control parameters. The figure presents $\langle\Phi_+|\hat{\overline{\mathscr{B}}}_\Phi\big(P{}_\mathrm{RTN}(\xi)\big)|\Phi_+\rangle$ as a function of $\omega_p$---the central frequency of the pulse-sequence-induced filter passband (it is related to the interpulse delay $\tau_p=\pi/\omega_p$). The Gaussian noise threshold $\overline{\mathscr{B}}_0\approx 2.19$ is indicated by the horizontal dashed line. The criterion is non-Gaussian positive if the curve passes over the threshold; the range of $\omega_p$ for which this is the case is indicated by the under-curve shading. Presented curves were obtained for different ratios of noise amplitude to the switching rate, $v/\gamma$. The upper and lower figures present the case of $n=2$ and $n=4$ pulses respectively, which results in the total duration of the evolution $T=n\pi/\omega_p$. For longer durations (greater number of pulses) the dephasing is generally stronger, and the resulting decoherence causes more sever damage to the state, which in turn lowers the effectiveness of the criterion.}
\label{fig:non-gauss_crit}
\end{figure}

Below we present an example of a physically realizable system, that would allow for implementation of the separability test \eqref{eq:avg_sep_crit} utilizing only standard experimental techniques. Spin qubits based on nitrogen-vacancy (NV) centers in diamond are currently a subject of intense experimental research aimed at using them as sensors of magnetic fields generated by single molecules \cite{Staudacher_Science13,DeVience_NN15,Haberle_NN15,Wrachtrup_JMR16,Lovchinsky_Science16,Degen_RMP17}. For a molecule consisting of a large number of atoms, the noisy magnetic field generated by nuclei of these atoms can be approximately treated as Gaussian noise \cite{Szankowski_JPCM17} with spectral denisty consisting of narrow peaks centered at frequencies of Larmor precession of distinct nuclear species. NV centers subjected to dynamical decoupling sequences with $\omega_p$ tuned to these Larmor frequencies of a given nuclear population, have been used to sense single molecules \cite{Lovchinsky_Science16}. A system consisting of two NV center qubits localized in the vicinity of a single molecule is thus a good candidate for demonstration of the above-discussed protocol. As it was discussed in Ref.~\cite{Krzywda_PRA17}, where interaction of two NV centers with such a noise source was considered, for a particular arrangement of qubits and molecule positions, the directions of qubit's quantization axes, and the axis of nuclear spin precession induced by the external magnetic field, it is possible to achieve perfect correlation of noises experienced by the two qubits.
With such a setup, the attenuation functions are given by $\chi = 2T^2 g^2/\pi$, 
where $g$ is a dipole-dipole coupling for the arrangement that gives perfect correlation \cite{Krzywda_PRA17}. 
Taking $T\approx g^{-1}\sqrt{\pi \ln(2)/6}$ we obtain the value of $\chi$ required to achieve the maximum of \eqref{eq:bound}.

\section{Criterion for detection of non-Gaussian statistics of the noise}\label{sec.crit}

The maximal value $\overline{\mathscr B}_0=1+2^{\frac{2}{3}}-2^{-\frac{4}{3}}\approx 2.19$ [see Eq.~\eqref{eq:bound}] obtained for Bell states in the case of perfectly correlated Gaussian noise, is in fact also the maximal value attainable for noises with Gaussian statistics in general (see Appendix \ref{app:max_B_gauss}). However, it is not the overall maximal value possible for $|\langle\mathcal{S}_\pm|\hat{\overline{\mathscr{B}}}_\mathcal{S}|\mathcal{S}_\pm\rangle|$ (with $\mathcal{S}=\Phi,\Psi$). Therefore, assuming that the noise field $\xi(\mathbf{r},t)$ coupled to the qubits has zero average and is stationary, the separability criterion \eqref{eq:avg_sep_crit} can be repurposed as a criterion for discriminating noises with non-Gaussian statistics:
\begin{equation}
\label{eq:non-gauss_crit}
\left(\begin{array}{c}
\text{If }|\langle\mathcal{S}_\pm|\hat{\overline{\mathscr{B}}}_\mathcal{S}\big(P(\xi)\big)|\mathcal{S}_\pm\rangle| > 
	1+2^{\frac{2}{3}}-2^{-\frac{4}{3}}\approx 2.19,\\[.3cm]
\text{then $P(\xi)$ is non-Gaussian.}\\
\end{array}\right),
\end{equation}
where we have included an argument for \CHSHlike{} operator to reiterate that it depends on the probability distribution of the noise $P(\xi)$.

An example of positive test by the criterion can be demonstrated with perfectly correlated random telegraph noise, a non-Gaussian stochastic process that jumps between two values, $v$ and $-v$, at the average rate $\gamma=(2\tau_c)^{-1}$, where $\tau_c$ is the correlation time of the noise \cite{Galperin_03,Szankowski2013,Ramon_PRB2015}. Its spectral density is a Lorentzian of width $2\pi \gamma$ centered at the zero frequency, hence the ``$\off{}$'' is realized by choosing $\omega_p \gg 2\pi \gamma=\pi/\tau_c$. Figure \ref{fig:non-gauss_crit} depicts the capabilities of criterion \eqref{eq:non-gauss_crit} to detect non-Gaussianity of the noise, depending on the ratio $v/\gamma$, and the choice of $\omega_p$ and the width of the passband (measured in the inverse of the number of pulses, or equivalently, in $T^{-1}$) of the ``$\on{}$'' control setting. The analytical results presented in the figure were obtained using the expressions for phase evolution of dynamically decoupled qubit coupled to random telegraph noise \cite{Ramon_PRB2015} adapted to the case of two qubits coupled to perfectly correlated noises. For more details see Appendix \ref{app:RTN}.

\section{Conclusions}\label{sec.conc}

We have shown that the coupling to a environment can be used as a tool for the detection of the entanglement between two qubits.
In our approach, the correlated sources of decoherence are a part of the mechanism triggering the local operations necessary to construct the separability criterion. The other part is the pulse-control method, which allows to locally fine-tune the strengths of the qubit-noise coupling. 
The correlators required for constriction of CHSH inequality are obtained by performing a measurement of fixed spin projections, after the duration of noise-driven evolution, during which the qubits are controlled with distinct choices of pulse sequence settings. We have shown that the inequality obtained in this procedure is a true criterion for entanglement---i.e, it is violated only by non-separable initial states.
Finally, we have also demonstrated that the level of the violation of the \CHSHlike{} inequality might provide the information whether the noise had Gaussian statistics or not.

Let us finish with one more remark regarding relation between the above-discussed scheme, that can be thought of as CHSH inequality averaged over an ensemble of measurement settings, 
and considerations on relation between violation of Bell inequalities and non-locality. 
As we have discussed at length, the {\it correlation} between the noises affecting the two qubits, 
that is equivalent to correlation between random measurement settings for the two qubits, is a crucial part of the proposed protocol. Such an explicit creation of correlations between measurement settings 
amounts to a violation of ``free choice'' assumption that is necessary for relatively straightforward establishment of relation between violation of Bell inequality and ruling out various kinds of 
local hidden variable models. While partial breaking of this assumption still allows for detection of quantum non-locality \cite{Putz_PRL14}, subtle considerations of this issue are beyond the scope of 
this work, in which we simply focused on decoherence-activated detection of entanglement.

\section*{Acknowledgements}
This work is supported by funds of Polish National Science Center (NCN), grant no.~DEC-2015/19/B/ST3/03152. We woud like to thank J.~Kaniewski, R.~Augusiak, and R.~Demkowicz-Dobrza{\'n}ski for enlightening discussions and comments.

\appendix

\begin{widetext}
\section{The noise-averaged correlator}\label{app:avg_corr}
Here we demonstrate how \eqref{eq:av_corr_op} has been obtained, as an average over realizations of noise $\xi$ of the respective correlator operator. Namely, we have
\begin{align}
  \nonumber
  \hat{\overline{E}}(a,b) &=\ 
  \int\mathcal{D}\xi P(\xi) \,\hat U_\xi^\dagger \hat\sigma_x^{(A)}\otimes\hat\sigma_x^{(B)}\hat U_\xi\\
  \nonumber
  &=\ \int\mathcal{D}\xi P(\xi) \,
  \left( \cos\alpha_\xi(a)\hat\sigma_x^{(A)}+\sin\alpha_\xi(a)\hat\sigma_y^{(A)}\right)\otimes\left( \cos\beta_\xi(b)\hat\sigma_x^{(B)}+\sin\beta_\xi(b)\hat\sigma_y^{(B)}\right)\\
  \nonumber
  &=\ \frac{\hat\sigma_x^{(A)}\otimes\hat\sigma_x^{(B)}}{2}\mathrm{Re}\int\mathcal{D}\xi P(\xi) \left( e^{i(\alpha_\xi(a)+\beta_\xi(b))}+e^{i(\alpha_\xi(a)-\beta_\xi(b))}\right)\\
  \nonumber
  &\ +\frac{\hat\sigma_y^{(A)}\otimes\hat\sigma_y^{(B)}}{2}\mathrm{Re}\int\mathcal{D}\xi P(\xi) \left( e^{i(\alpha_\xi(a)+\beta_\xi(b))}-e^{i(\alpha_\xi(a)-\beta_\xi(b))}\right)\\
  \nonumber
  &\ +\frac{\hat\sigma_x^{(A)}\otimes\hat\sigma_y^{(B)}}{2}\mathrm{Im}\int\mathcal{D}\xi P(\xi) \left( e^{i(\alpha_\xi(a)+\beta_\xi(b))}+e^{i(\alpha_\xi(a)-\beta_\xi(b))}\right)\\
  \nonumber
  &\ +\frac{\hat\sigma_y^{(A)}\otimes\hat\sigma_x^{(B)}}{2}\mathrm{Im}\int\mathcal{D}\xi P(\xi) \left( e^{i(\alpha_\xi(a)+\beta_\xi(b))}-e^{i(\alpha_\xi(a)-\beta_\xi(b))}\right)\\
  &=\ \mathrm{Re}\big\{\phi[\,\alpha_\xi(a)+\beta_\xi(b)\,]\big\}\frac{\hat{E}(0,0)+\hat{E}(\tfrac{\pi}{2},\tfrac{\pi}{2})}{2}
  +\mathrm{Re}\big\{\phi[\,\alpha_\xi(a)-\beta_\xi(b)\,]\big\}\frac{\hat{E}(0,0)-\hat{E}(\tfrac{\pi}{2},\tfrac{\pi}{2})}{2}
\end{align}
\end{widetext}
Here, $\hat{E}(\theta_A,\theta_B)$ are given by \eqref{eq:corr_op}, and the exponential functions averaged over noise realizations were identified with the characteristic functions of stochastic phases $\alpha_\xi(a)\pm\beta_\xi(b)$,
\begin{equation}\label{eq:app:char_func}
\phi[\theta_\xi ]=\int\mathcal{D}\xi P(\xi)\,e^{i\theta_\xi}\,,
\end{equation}
and we assume that characteristic functions are purely real. The logarithm of characteristic function is the cumulant generating function $\chi$ of the stochastic phase,
\begin{equation}\label{eq:app:cumulant_gen}
\chi[\theta_\xi] = \ln \phi[\theta_\xi]\,,
\end{equation}
that defines its cumulants (i.e., the correlation functions), according to
\begin{equation}
\kappa_k[\theta_\xi] = \frac{1}{k!}\frac{\partial^k}{\partial v^k}\chi[ v\theta_\xi ]\Big|_{v=0}.
\end{equation}
The local and non-local attenuation functions are thus obtained as
\begin{subequations}
\begin{align}
\chi_A(a) &\equiv \sum_{k=1}^\infty \kappa_{2k}[\alpha_\xi(a)]\,,\\
\chi_B(b) &\equiv \sum_{k=1}^\infty \kappa_{2k}[\beta_\xi(a)]\,,\\
\chi_{\{AB\}}(a,b) &\equiv \chi[\alpha_\xi(a)+\beta_\xi(b)]-\chi_{A}(a)-\chi_{B}(b),\\[.2cm]
\chi_{[AB]}(a,b) &\equiv \chi[\alpha_\xi(a)-\beta_\xi(b) ]-\chi_A(a)-\chi_B(b).
\end{align}
\end{subequations}

\section{The fundamental properties of $\hat{\overline{\mathscr{B}}}_\mathcal{S}$}\label{app:prop_avg_CHSH_op}

(i) Formally, operators $\hat{\overline{\mathscr{B}}}_\mathcal{S}$ are given by the standard CHSH operator average over noise realizations. Therefore, one can write
\begin{align}
\nonumber
&|\mathrm{Tr}\,\hat{\overline{\mathscr{B}}}_{\Phi/\Psi} \hat\varrho_\mathrm{sep} |=\\
\nonumber
&= \left|\int\mathcal{D}\xi P(\xi)\mathrm{Tr}\,\hat{\mathscr{B}}_{\{ \alpha_\xi(\off{}),\alpha_\xi(\on{}),\beta_\xi(\off{}),\pm\beta_\xi(\on{})\}}
	 	\hat\varrho_\mathrm{sep}\right|\\
\nonumber
&\leqslant  \int\mathcal{D}\xi P(\xi)|\mathrm{Tr}\,\hat{\mathscr{B}}_{\{ \alpha_\xi(\off{}),\alpha_\xi(\on{}),\beta_\xi(\off{}),\pm\beta_\xi(\on{})\}}
	 	\hat\varrho_\mathrm{sep}|\\
&\leqslant \int\mathcal{D}\xi P(\xi)\, 2 = 2\,.
\end{align}
Note that using similar reasoning one can show that the maximal value of the expectation value on arbitrary state is $2\sqrt{2}$---the same value as for standard CHSH operator.

(ii) Assume that $\chi_{\{AB\}}=\chi_{[AB]}=0$, and suppose that $|\mathrm{Tr}\,\hat{\overline{\mathscr{B}}}_\mathcal{S}\hat\varrho|>2$ (note that for uncorrelated noises both \CHSHlike{} operators are identical), then
\begin{align*}
&|\mathrm{Tr}\,\hat E(0,0)\hat\varrho|(1+e^{-\chi_A}+e^{-\chi_B}-e^{-\chi_A-\chi_B}) >2 \,\, , \\[.3cm]
&e^{-\frac{\chi_A+\chi_B}2}\left(e^{\frac{\chi_B-\chi_A}2}+e^{-\frac{\chi_B-\chi_A}2}-e^{-\frac{\chi_A+\chi_B}2}\right)>1 \,\, ,\\
&2\cosh\left(\frac{\chi_B-\chi_A}2\right)> 2\cosh\left(\frac{\chi_A+\chi_B}2\right)\,.
\end{align*}
 Since $\chi_Q\geqslant 0$ and $\cosh$ is a monotonic function, we have arrived at a contradiction. Therefore, $|\mathrm{Tr}\,\hat{\overline{\mathscr B}}_\mathcal{S}\hat\varrho|\leqslant 2$ for uncorrelated noises.
 
\section{The maximal value of $|\langle\mathcal{S}_\pm|\hat{\overline{\mathscr{B}}}_{\mathcal{S}}|\mathcal{S}_\pm\rangle|$ for the case of Gaussian noise}\label{app:max_B_gauss}

From the structure of \eqref{eq:avg_CHSH_op} it is evident that $|\langle\mathcal{S}_\pm|\hat{\overline{\mathscr{B}}}_\mathcal{S}|\mathcal{S}_\pm\rangle|$ (with $\mathcal{S}=\Phi,\Psi$) can be made larger when $|\langle\mathcal{S}_\pm|\hat{\overline{E}}(\on{},\on{})|\mathcal{S}_\pm\rangle|$ is made smaller. For fixed $\chi_Q(\on{})$, if one assumes Gaussian statistics of the noise [i.e., that the attenuation functions are given by \eqref{eq:gauss_att_func}], then this correlator can be made smaller by making $\chi_{AB}(\on{},\on{})$ as large as possible. For Gaussian noise, the non-local attenuation function satisfy Cauchy-Schwartz inequality $|\chi_{AB}(\on{},\on{})|\leqslant\sqrt{\chi_A(\on{})}\sqrt{\chi_B(\on{})}$, and one instance when this inequality is saturated is for perfectly correlated noises, when $\chi_A(\on{})=\chi_B(\on{})$. Hence, perfect correlation is sufficient to obtain the maximal value of $|\langle\mathcal{S}_\pm|\hat{\overline{\mathscr{B}}}_\mathcal{S}|\mathcal{S}_\pm\rangle|$, because it gives the maximal damping of  $|\langle\mathcal{S}_\pm|\hat{\overline{E}}(\on{},\on{})|\mathcal{S}_\pm\rangle|$, and in addition it results in \eqref{eq:avg_CHSH_Gauss}.

\begin{widetext}
\section{Noise-average CHSH operator for perfectly correlated random telegraph noises}\label{app:RTN}

In Ref.~\cite{Ramon_PRB2015} it was shown that the characteristic function \eqref{eq:app:char_func} of stochastic phase $\theta_\xi=\int_0^T dt f(t)\xi(t)$, where $\xi(t)$ is a random telegraph noise with amplitude $v$ and the switching rate $\gamma$, and $f(t)$ is the time-domain filter function of Carr-Purcell sequence with $n$ pulses and $\tau_p = \pi/\omega_p$ interpulse interval, is given by
\begin{align}
  \phi[\theta_\xi] = W(v,n,\gamma,\tau_p)=\frac{e^{-\gamma n \tau_p}}{2\mu^n}
  \left(\frac{\cosh(\gamma\mu\tau_p)-v^2/\gamma^2}{\mu\sqrt{\sinh^2(\gamma\mu\tau_p)+\mu^2}}(\lambda_+^n-\lambda_-^n)
  +(\lambda_+^n+\lambda_-^n)\right)\,,
\end{align}
where
\begin{align}
  \mu &= \sqrt{1-\frac{v^2}{\gamma^2}}\,,\\
  \lambda_\pm &= \sinh(\gamma\mu\tau_p) \pm \sqrt{\sinh^2(\gamma\mu\tau_p)+\mu^2}\,.
\end{align}
\end{widetext}
The characteristic function is related to the cumulant generating functional (and to attenuation functions) via Eq.~\eqref{eq:app:cumulant_gen}. For perfectly correlated noises we have
\begin{equation}
\chi_A+\chi_B+2\chi_{\{AB\}} =\chi[\alpha_\xi+\beta_\xi] =\chi[2\alpha_\xi]=\chi[\alpha_{2\xi}]\,,
\end{equation}
that is, the cumulant generating functional of a sum of stochastic phases is identical to $\chi$ of stochastic phase acquired by coupling to a single noise with twice the amplitude. Therefore we can write
\begin{align}
  e^{-\chi_Q} &= W(v,n,\gamma,\tau_p)\,,\\
  e^{-\chi_A-\chi_B-2\chi_{\{AB\}}} &=W(2v,n,\gamma,\tau_p)\,,
\end{align}
and thus we obtain the analytical expression for the expectation value of \CHSHlike{} operator
\begin{equation}
\langle\Phi_+|\hat{\overline{\mathscr{B}}}_\Phi|\Phi_+\rangle
	=1+2W(v,n,\gamma,\tau_p)-W(2v,n,\gamma,\tau_p)\,.
\end{equation}


%

\end{document}